\documentclass{article}

\usepackage{arxiv}

\usepackage{cite}
\usepackage{amsmath,amssymb,amsfonts}
\usepackage{algorithmic}
\usepackage{graphicx}
\usepackage{textcomp}
\usepackage{xcolor}
\def\BibTeX{{\rm B\kern-.05em{\sc i\kern-.025em b}\kern-.08em
    T\kern-.1667em\lower.7ex\hbox{E}\kern-.125emX}}

\usepackage{multirow}
\usepackage{adjustbox}
\usepackage{float} 
\usepackage{wrapfig}
\usepackage{soul}
\usepackage{color}
\usepackage{tabularx}
\usepackage{cleveref}

 \makeatletter
    \newcommand{\linebreakand}{%
      \end{@IEEEauthorhalign}
      \hfill\mbox{}\par
      \mbox{}\hfill\begin{@IEEEauthorhalign}
    }
    \makeatother

\crefname{table}{Table}{Tables}
\Crefname{table}{Table}{Tables}

\begin{document}

\title{User Experience Evaluation of AR Assisted Industrial Maintenance and Support Applications}

\author{Akos Nagy \\
Department of Networks \& Digital Media \\
School of Computer Science \& Maths,\\
ECE \\
Kingston University\\
Kingston upon Thames, UK\\
\textit{A.Nagy@kingston.ac.uk} \\
\And
Yannis Spyridis \\
Department of Networks \& Digital Media  \\
School of Computer Science \& Maths, \\
ECE\\
Kingston University\\
Kingston upon Thames, UK \\
\textit{Y.Spyridis@kingston.ac.uk} \\
\And
Gregory J Mills\\
Department of Networks \& Digital Media  \\
School of Computer Science \& Maths, \\
ECE\\
Kingston University\\
Kingston upon Thames, UK \\
\textit{G.Mills@kingston.ac.uk} \\
\And
Vasileios Argyriou \\
Department of Networks \& Digital Media  \\
School of Computer Science \& Maths, \\
ECE \\
Kingston University\\
Kingston upon Thames, UK \\
\textit{Vasileios.Argyriou@kingston.ac.uk} \\
}

\maketitle

\begin{abstract}
The paper introduces an innovative approach to industrial maintenance leveraging augmented reality (AR) technology, focusing on enhancing the user experience and efficiency. The shift from traditional to proactive maintenance strategies underscores the significance of maintenance in industrial systems. The proposed solution integrates AR interfaces, particularly through Head-Mounted Display (HMD) devices, to provide expert personnel-aided decision support for maintenance technicians, with the association of Artificial Intelligence (AI) solutions. The study explores the user experience aspect of AR interfaces in a simulated industrial environment, aiming to improve the maintenance processes' intuitiveness and effectiveness. Evaluation metrics such as the NASA Task Load Index (NASA-TLX) and the System Usability Scale (SUS) are employed to assess the usability, performance, and workload implications of the AR maintenance system. Additionally, the paper discusses the technical implementation, methodology, and results of experiments conducted to evaluate the effectiveness of the proposed solution.
\end{abstract}

\keywords{Augmented Reality, HCI, Industrial Maintenance, User Experience}

\section{Introduction}
The manufacturing industry has evolved over the years, affecting various departments, with maintenance being a key area of significant change. Maintenance is vital for ensuring production efficiency, as unexpected disruptions can lead to decreased system performance, productivity loss, and missed business opportunities \cite{MOURTZIS2016655}. Traditionally, maintenance strategies were mainly reactive and preventive, with predictive approaches applied in critical situations. However, the maintenance paradigm is shifting, recognizing its importance as a strategic factor and profit contributor for industrial systems.

The proposed solution aims to create an intuitive and modular maintenance system based on Industry 5.0 principles\cite{en15145221, asi5010027}. Maintenance processes are typically carried out by trained personnel following established procedures. The application of augmented reality (AR) technologies can assist maintenance technicians by providing guided decision support through human-machine interaction.

The usage of AR technology \cite{PALMARINI2018215, ESWARAN2023118983, DANIELSSON2020100175, 10.1145/2684103.2684135}, which superimposes virtual entities onto the real world, including images, texts, audiovisuals, and 2D/3D models, improves complex maintenance tasks, by providing additional information for the maintenance personnel. Our paper explores the user experience aspect of an AR interface in guiding maintenance personnel, employing Head-Mounted Display (HMD) devices \cite{10.1145/3282894.3289745}, and a support interface for expert personnel, in a simulated industrial environment. The interfaces provide communication between personnel and the utilization of Artificial Intelligence (AI) solutions in the form of Computer Vision models. \cite{nalbant2021computervision, 10110300}

\section{Related Work}

The user interface (UI) system, evaluated by Ana et al.\cite{8927815}, is designed to guide technicians through maintenance tasks using AR. It positions the technician at the procedure location and provides step-by-step guidance with media, such as images and 3D models, to explain the task. The system uses both a head-mounted display (Microsoft HoloLens) and a mobile Android tablet (Lenovo 10") for hardware support. The software tools used include the Vuforia AR SDK for marker-based applications and Unity3D for development. The design requirements emphasize a wide field of view, lightweight AR interfaces, long-lasting batteries, optical and retinal projection, and voice-based interaction. The UI system architecture supports training and execution of maintenance tasks, maintaining a list for operators to track their progress. It also offers access to the history of completed maintenance procedures.

The visualization design includes support for training and execution modes, allowing users to practice and follow steps on the mobile device before executing the task in real-time using AR. The system keeps a history of completed procedures, making it easy for users to review and re-practice tasks. There is also support for technical assistance when needed.

Implementation involves a main menu with options for training, maintenance, and monitoring. In training and maintenance modes, the system provides step-by-step guides with various media formats to support operators during their tasks. 3D animations and support/help features are used to guide users and address problems. Monitoring mode provides data about machine conditions. Users can access a history window to view their completed procedures and task durations.

The applicability of a similar framework to ours, proposed by Dimitris et al.\cite{app10051855}, was tested and validated both in a laboratory-based machine shop and in a real-life industrial scenario. The industrial scenario involved the use of machine tools in an existing machine shop, which had machines from different technological eras. This diversity made it important to assess the framework's impact on varying machine tools.

The solution consists of a technician wearing a HoloLens HMD to perform maintenance operations, guided by an expert engineer in the background to communicate with the technician. The practical implementation and validation of the framework happened in a laboratory-based scenario. An expert engineer, from their desktop, views the technician's field of view (FoV) containing the malfunctioning machine tool. The Microsoft HoloLens's spatial recognition algorithm recognizes the technician's physical environment, allowing the engineer to overlay 3D augmentations on the user's surroundings. A shop-floor technician establishes a video teleconference with an engineer to receive guidance, with a communication window opening in the technician's FoV for audio and video communication.

\section{Methodology}

This section will discuss the developed applications, procedures, and evaluation metrics involved in the evaluation process.

\subsection{AR Maintenance Application}

\begin{figure}[h]
    \centering
    \includegraphics[width=0.8\textwidth]{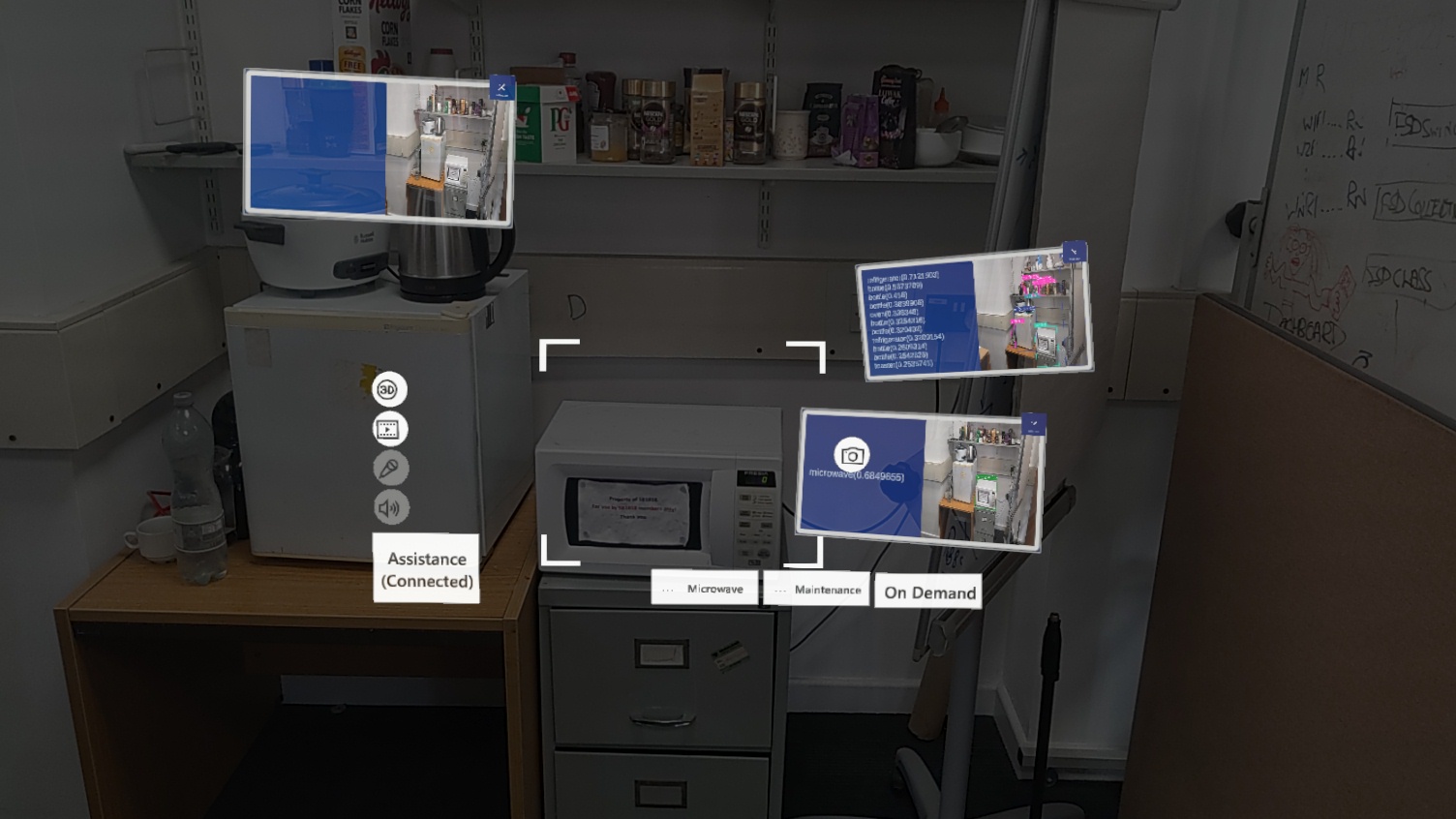}        
    \caption{AR Application - Displaying AI scene analysis results on virtual cards}
    \label{fig:AR_AI_Results}
    \vspace{0pt}
\end{figure}

The AR Maintenance application was developed to provide information for the maintenance personnel during maintenance processes and was deployed on an HMD device. The main functionality of the AR application is to display data for the maintenance personnel based on the work environment.
The application allows the user to take a picture using the camera of the HMD device. AI models then process the image using computer vision techniques to identify target objects within the environment. The maintenance personnel are given the option to select which AI model needs to be involved in the process, optimizing the results to the specific need, while reducing the computational needs of the request. The general user interface and the visualized results panel, displaying the taken image and the identified objects, are shown on \cref{fig:AR_AI_Results}.

\begin{figure}[h]
    \centering
    \includegraphics[width=0.8\textwidth]{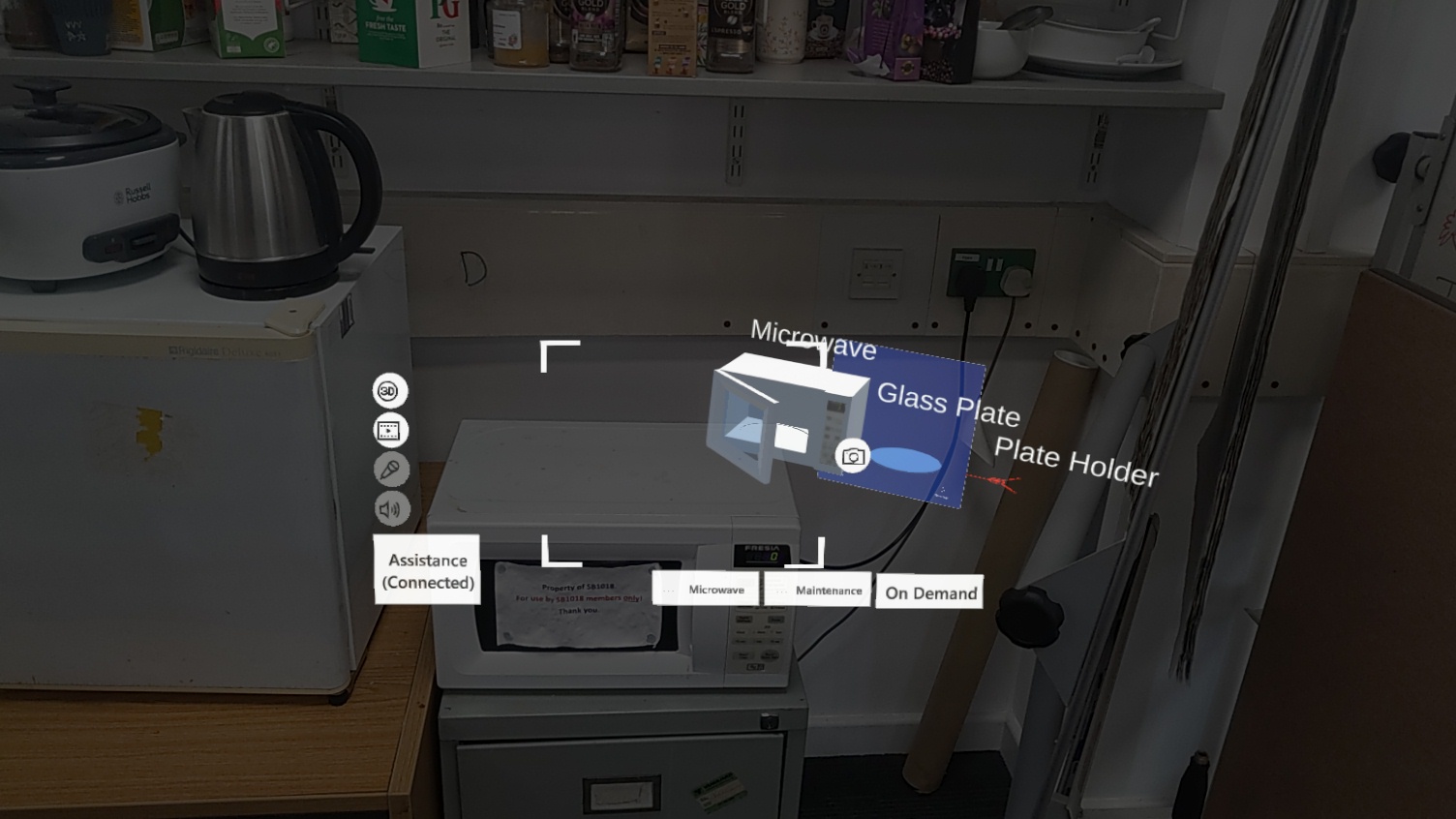}        
    \caption{AR Application - Utilizing Action Panels to display information (3D object)}
    \label{fig:AR_Action_Panel}
    \vspace{0pt}
\end{figure}

The secondary functionality of the AR application is to provide communication with expert support personnel via a multitude of communication channels. The voice channel allows verbal communication, while the video channel, utilizing the camera sensor of the HMD device, broadcasts the view of the maintenance personnel, including the virtual objects visualized by the device. Utilizing the spatial awareness feature of the HMD device, which creates a 3D approximation of the work environment, the application provides additional information about the environment.
Utilizing the AR functionalities, that allow 3D interactive virtual objects to be displayed for the maintenance personnel, the application can receive digital instructions from the support personnel. The "Action Cards", as seen in \cref{fig:AR_Action_Panel}, displays text-, image-, video-, or 3D object-based information.

\subsection{Support Application}

The support application was developed to provide communication with the maintenance personnel. The application utilizes the same communication channels as the AR application, providing real-time verbal communication, while a visual feed displaying both the work environment and the virtual user interface projected by the AR device. 

\begin{figure}[h]
    \centering
    \includegraphics[width=0.8\textwidth]{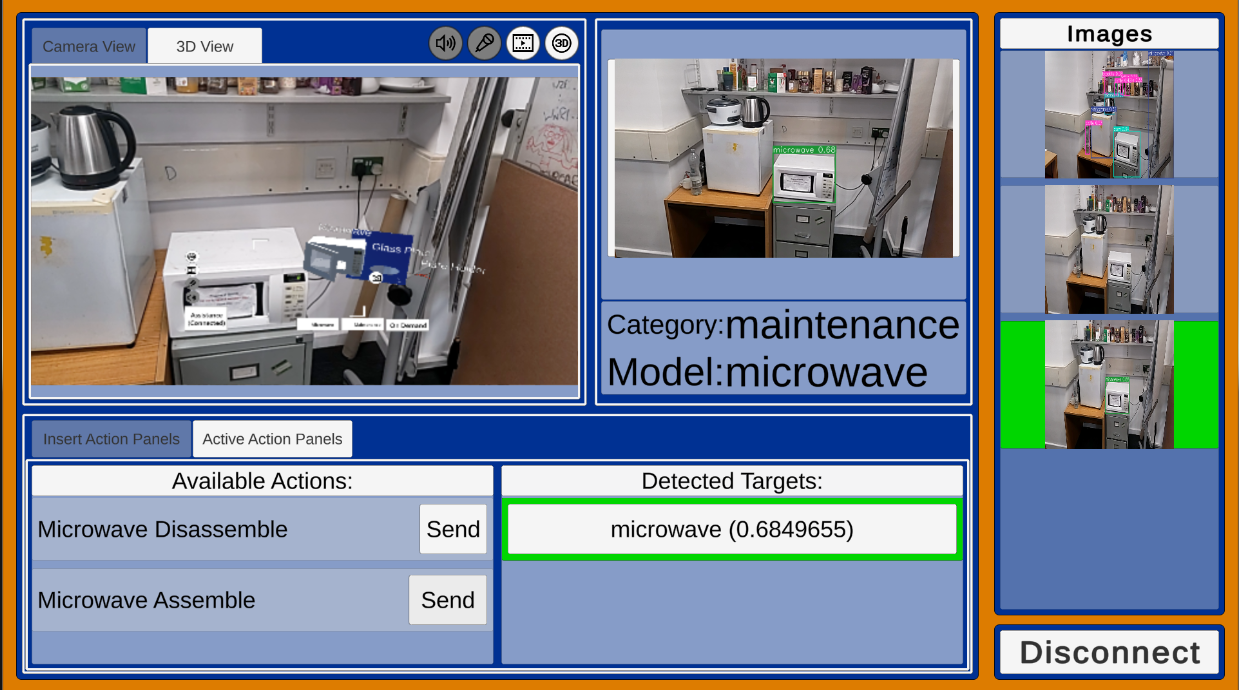}        
    \caption{Support Application - Main user interface}
    \label{fig:SUPPORT_UI}
    \vspace{0pt}
\end{figure}

While the maintenance and support personnel are connected, all images taken by the maintenance personnel and processed by a computer vision algorithm are available for the support personnel, providing the support personnel with the results of the image processing, as seen in \cref{fig:SUPPORT_UI}. In addition, to the information provided to the support personnel, the application also allows the support personnel to provide "Action Cards", discussed in the previous section, that will show information or instructions for the maintenance personnel, which are related to certain detected targets. Furthermore, the support application allows the personnel to change the information displayed on the "Action card" in case the maintenance personnel not able to interact with the virtual objects due to the complexity of the performed maintenance task.

\begin{figure}[h]
    \centering
    \includegraphics[width=0.8\textwidth]{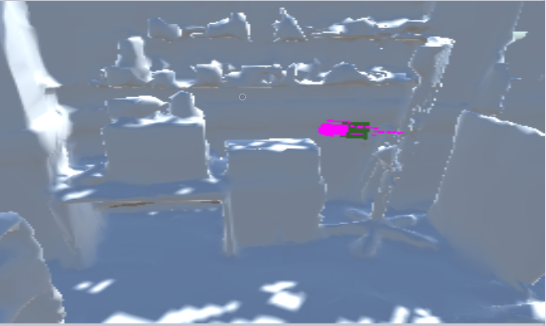}        
    \caption{Support Application - 3D visualization of the work environment for the support personnel}
    \label{fig:SUPPORT_3D_View}
    \vspace{0pt}
\end{figure}

Provided by the spatial awareness feature of the HMD device, the scanned work environment can also be displayed for the support personnel. Although the 3D recreation of the environment does not contain color data \cite{s20041021}, and only being presented via a single colored 3D object, as seen in \cref{fig:SUPPORT_3D_View}, it allows the support personnel to have an approximation of the work environment. The application also displays the position and orientation of the maintenance personnel in the 3D environment, alongside the 3D interactive objects presented to the maintenance personnel via the AR device. A secondary feature of the 3D environment viewer interface is that it allows the support personnel to point at a place in the real world, creating an indicator, that can further assist the maintenance personnel in orienting in the work environment.

\subsection{Evaluation Procedure}
The evaluation methodology is divided into experiments, where each experiment requires two participants. The participants assume the role of in-training maintenance personnel and in-training support personnel. The maintenance personnel situated in the maintenance area, are equipped with an AR HMD device, while the Support Personnel use a personal computer in a separate room. 

The goal of the participants is to identify, localize, and solve issues with a faulty machine. The experiment happens in a simulated environment where the participants do not require prior knowledge of the involved technologies or the inner workings of the involved machines. The tasks are performed in a safe environment, where the involved machines are common kitchen appliances, and the activities do not require more complex actions than a normal day-to-day life would require. 

\begin{figure}
    \centering
    \includegraphics[width=0.8\textwidth]{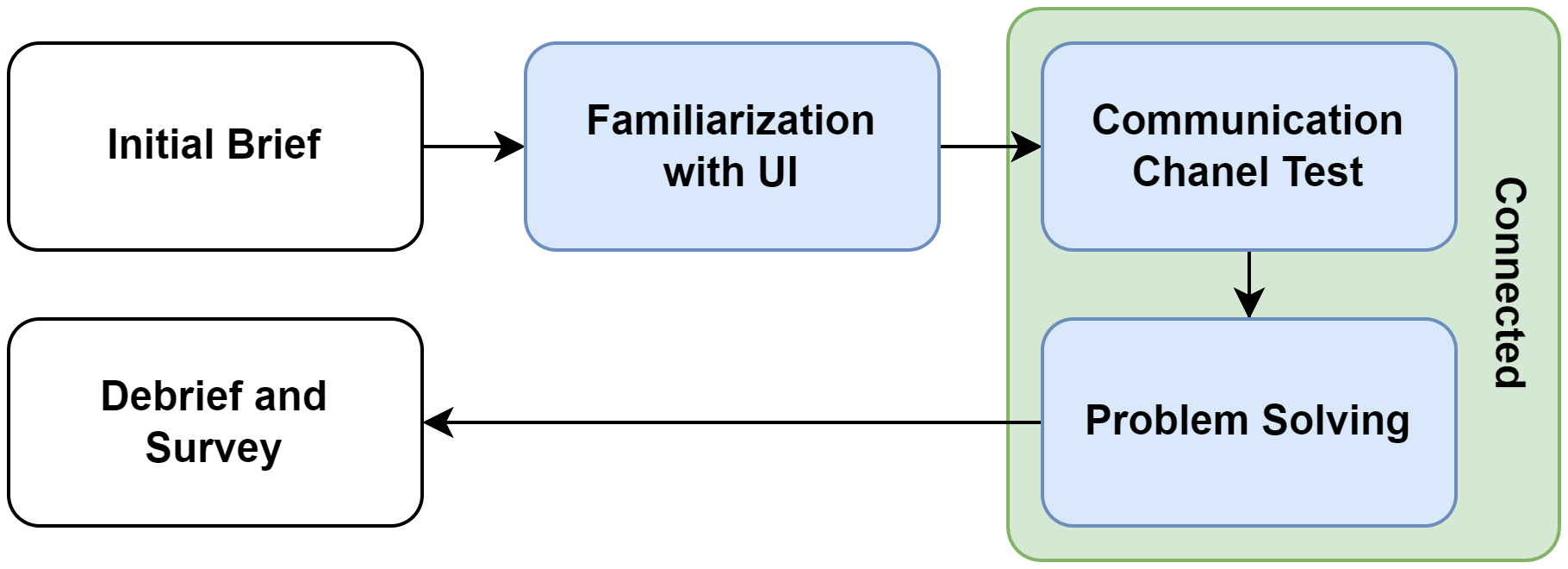}        
    \caption{Overview of evaluation workflow}
    \label{fig:Acrhitecture}
    \vspace{0pt}
\end{figure}

An experiment, as displayed in \cref{fig:Acrhitecture}, encompasses four key phases: Briefing, Familiarization, Maintenance, and Debriefing Phase and Qualitative Data Collection. 

\textbf{Briefing Phase}

This is the initial stage, where participants are comprehensively briefed about the experiment. They're provided with detailed information about the experiment's nature, the specific tasks they'll be undertaking, and the data that will be collected. Additionally, the briefing outlines the overarching premise and goals of the exercise, ensuring participants understand the context and objectives before accepting their participation and moving forward. 

\textbf{Familiarization Phase}

Here, participants are given the opportunity to familiarize themselves with the environment in which the experiment will take place. They are encouraged to explore and become acquainted with the technology and interfaces relevant to the experiment's goals. Special attention is directed towards ensuring participants grasp the capabilities of the AR technology and AR HMD device involved. This phase aims to equip participants with a level of familiarity essential for effective engagement during the experiment. 

\textbf{Maintenance Phase}

In this pivotal stage, support personnel are presented with scenarios that challenge their problem-solving abilities. The focus is on resolving issues related to a simulated dummy machine. Utilizing a meticulously designed protocol, participants are guided through step-by-step procedures mimicking real maintenance scenarios. Multiple features of the solution are employed, including voice communication channels, visual aids such as image feeds, the utilization of a 3D recreation of the maintenance environment, and the use of action panels to provide crucial information to aid the maintenance personnel. This phase is a practical application of skills and knowledge acquired during the familiarization phase. 

\textbf{Debriefing Phase and Qualitative Data Collection}

Concluding the experiment, the debriefing phase is essential for gathering comprehensive feedback and data. Participants are presented with tailored questionnaires designed to align with the specific data collection objectives of the experiment. These questionnaires are carefully crafted to capture relevant insights and perceptions from participants. Moreover, conversations are initiated to gauge participant satisfaction with the testing environment. These discussions serve the dual purpose of ensuring participant comfort and gathering any additional qualitative information that may not have been covered in the questionnaires. This phase provides valuable insights into participant experiences and perceptions, adding depth to the collected data. 

\subsection{Qualitative Metrics}

For the Qualitative Metrics, two standard evaluation methods were utilized, the NASA Task Load Index and the System Usability Scale, alongside additional data points that are specific to the two applications discussed in the paper.

The \textbf{NASA Task Load Index} (NASA-TLX) \cite{HART1988139, hart2006nasa} is a multidimensional assessment tool utilized within human factors research and ergonomics to gauge subjective workload perceptions experienced by individuals engaged in various tasks. Developed by Hart and Staveland in the 1980s, NASA-TLX has become a prominent instrument due to its capacity to quantify workload across diverse domains such as aviation, healthcare, and industrial settings.

The NASA-TLX can be scored using different versions, for the current paper the raw scoring method has been chosen. Users rate each of the six components on a scale from 0 to 100, with 0 representing the lowest level of demand or workload and 100 representing the highest level. After collecting ratings from users, you simply calculate the average of the six component scores to obtain the overall workload score. This method provides a straightforward and interpretable composite score.

The choice of scoring method may depend on the specific needs of the study or application. Raw scoring is the simplest and most straightforward, while weighted scoring and category scaling provide more nuanced insights into the nature of the workload. Researchers and practitioners can select the scoring method that best aligns with their objectives and the level of detail they require to assess and address workload issues in the task or system under consideration.

The \textbf{System Usability Scale} (SUS) \cite{doi:10.1080/10447310802205776, bangor} is a widely used questionnaire for assessing the perceived usability of a system, product, or service. It was developed by John Brooke in 1986 and has become a popular tool for gathering feedback and evaluating user experiences in various domains, including software applications, websites, and hardware devices. SUS is known for its simplicity and effectiveness in providing valuable insights into usability.

The SUS questionnaire consists of ten statements or questions that users rate on a scale from 1 to 5, with 1 representing "Strongly Disagree" and 5 representing "Strongly Agree". The questions are typically presented in a randomized order to minimize response bias. The ten statements are as follows:
\begin{itemize}
    \item I think that I would like to use this system frequently.
    \item I found the system unnecessarily complex.
    \item I thought the system was easy to use.
    \item I think that I would need the support of a technical person to be able to use this system.
    \item I found the various functions in this system were well integrated.
    \item I thought there was too much inconsistency in this system.
    \item I would imagine that most people would learn to use this system very quickly.
    \item I found the system very cumbersome to use.
    \item I felt very confident using the system.
    \item I needed to learn a lot of things before I could get going with this system.
\end{itemize}

Half of the statements were purposefully worded negatively, encouraging the participants to contemplate the individual statements. Due to this method, the given answers better represent the participants' opinions on the topic, while also allowing to filter false results where all the answers were either high or low values, indicating a non-responsible participation.

Alongside the standard evaluation methods use-case specific \textbf{additional qualitative data} has been collected. These statements focus on features of the applications such as the "3D representations" and "Action Cards" which are the novel features of these applications.
Participants were presented with statements and asked to evaluate their experiences on a 5-point Likert scale with 1 representing "Strongly Disagree" and 5 representing "Strongly Agree".

Regarding AR Application the following statements were presented to the participants:
\begin{itemize}
    \item I would consider using AR technology for similar scenarios
    \item I found the voice communication helpful
    \item I believe the 3D representation of the support personnel can be beneficial
    \item The ”Action cards” are useful for displaying information
    \item It’s beneficial that ”Action cards” could be controlled by the support personnel
    \item I would consider this method to have a direct connection with other people to improve productivity
\end{itemize}

While the participants using the Support Application were presented with the following statements:
\begin{itemize}
    \item I found the voice communication really helpful
    \item The ”Video Feed” made communication with the Maintenance personnel easier
    \item The ”3D View” helped with communication in the environment
    \item I would consider this method to have a direct connection with other people to improve productivity
\end{itemize}

\subsection{Techincal Implementation}

The current solution was developed using the Unity game engine developed by Unity Technologies. Communication between applications was achieved using PUN 2, a networking solution developed by Photon, and other, Python-based server solutions, utilizing Flask and Flask-SocketIO libraries. The AR Maintenance application was deployed on a HoloLens 2 HMD device developed by Microsoft.

\section{Results}

A total of 11 participants were involved in the experiments, 8 male and 3 female, with ages ranging from 18 to 44. The "Connected" segment of the experiment, which included the Communication Channel Tests and the Problem-solving phases, averaged 13 minutes and 24 seconds. 
Before this experiment, 63.6\% of the participants had experience using AR technologies in some capacity.

\subsection{System Usability Score Results}

When evaluating the raw usability score, we consider the midpoint of the scale (2.5) as an indicator of how the applications performed in a given category.

Based on the raw score data, as seen on \cref{table:SUSQR-AR}, the \textbf{Maintenance AR Application}, on average, can be considered to be moderately usable. 
The ease-of-use indicators have received a high average score of 4.1, indicating that the AR application, despite the lack of general familiarity with the AR technology, and specifically HMD devices, can be considered easy to use. Users felt confident using the application, with an average score of 4.3.

However, while participants perceived the various functions in the application to be well-integrated, with an average score of 4.0, there were some concerns regarding the consistency of the evaluated features, reflected in an average score of 2.4.

Participants generally felt that most people would learn to use the application quickly (average score of 4.3), despite having to learn some aspects before getting started (average score of 2.1). This is further supported by the relatively low average score of 2.2 regarding the need for support from a technical person while operating the application. The application received an even lower average score of 1.9 in the category of cumbersomeness, indicating that users did not find it overly cumbersome to use.

Participants expressed a moderate likelihood of using the application frequently, with an average score of 3.8.

\begin{table}[h]
\centering
\caption{System Usability Scale Questioner Results for the Maintenance AR Application}
\label{table:SUSQR-AR}
\begin{tabularx}{\columnwidth}{| X | c |}
\hline
\textbf{Question} & \textbf{Average Raw Value}\\
\hline
I think that I would like to use this system frequently. & 3.8 \\ 
\hline
I found the system unnecessarily complex. & 2.1 \\
\hline
I thought the system was easy to use. & 4.1 \\
\hline
I think that I would need the support of a technical person to be able to use this system. & 2.2 \\
\hline
I found the various functions in this system were well integrated. & 4.0 \\
\hline
I thought there was too much inconsistency in this system. & 2.4 \\
\hline
I would imagine that most people would learn to use this system very quickly. & 4.3 \\
\hline
I found the system very cumbersome to use. & 1.9 \\
\hline
I felt very confident using the system. & 4.3 \\
\hline
I needed to learn a lot of things before I could get going with this system. & 2.1 \\
\hline
\end{tabularx}
\end{table}

Users expressed a high likelihood of using the system frequently, with an average score of 4.0. The system was perceived as easy to use, scoring an average of 4.1. Additionally, users found the various functions in the system to be well-integrated, with a high average score of 4.3. Furthermore, users felt very confident while using the system, as indicated by a high average score of 4.4. They also believed that most people would quickly learn to use the system, scoring it with an average of 4.5, despite indicating a need to learn some things before getting started (average score of 2.0).

Although some users expressed a moderate need for technical support, with an average score of 2.5, it was not considered a major barrier to use. Users perceived very little inconsistency in the system, with an average score of 1.7. Additionally, the system received a relatively low rating for cumbersomeness, averaging 2.1, indicating that users did not find it overly cumbersome to use. These results are summarized in \cref{table:SUSQR-SUPPORT}. 

\begin{table}[h]
\centering
\caption{System Usability Scale Questioner Results for the Support Application}
\label{table:SUSQR-SUPPORT}
\begin{tabularx}{\columnwidth}{| X | c |}
\hline
\textbf{Question} & \textbf{Average Raw Value}\\
\hline
I think that I would like to use this system frequently. & 4 \\ 
\hline
I found the system unnecessarily complex. & 2.2 \\
\hline
I thought the system was easy to use. & 4.1 \\
\hline
I think that I would need the support of a technical person to be able to use this system. & 2.5 \\
\hline
I found the various functions in this system were well integrated. & 4.3 \\
\hline
I thought there was too much inconsistency in this system. & 1.7 \\
\hline
I would imagine that most people would learn to use this system very quickly. & 4.5 \\
\hline
I found the system very cumbersome to use. & 2.1 \\
\hline
I felt very confident using the system. & 4.4 \\
\hline
I needed to learn a lot of things before I could get going with this system. & 2 \\
\hline
\end{tabularx}
\end{table}

In comparing the average scores for "needing support from a technical person" between the AR Application (2.2) and the Support Application (2.5), it becomes apparent that participants, despite potentially being less familiar with AR technology, found the user interface of the AR Application more intuitively designed, thus requiring less assistance. The slightly higher score for the Support Application could be attributed to its broader scope of functionalities, necessitating a higher likelihood of seeking technical support.

Both sets of results also reveal similar tendencies regarding users' desire to frequently use the system, with scores of 3.8 for the AR Application and 4.0 for the Support Application. Additionally, users across both applications perceived the systems as easy to use, as evidenced by average scores of 4.1.

While there are minor discrepancies in perceived technical support needs, consistency, and learning curves, both datasets consistently indicate similar user perceptions regarding the frequency of usage, ease of use, integration of functions, cumbersomeness, and confidence in system operation. Consequently, both applications can be considered moderately usable, with users generally reporting positive experiences.

Overall, these findings suggest that despite differences in application domains and functionalities, both the AR Application and the Support Application offer satisfactory usability levels and contribute to positive user experiences.

\begin{table}[h]
\centering
\caption{Final System Usability Scale Values for the Support and Maintenance AR Application}
\label{table:SUSQR-OVERALL}
\begin{tabularx}{\columnwidth}{| X | c | c |}
\hline
\textbf{Application} & \textbf{SUS Score} & \textbf{Workload} \\
\hline
Maintenance Application & 74.54 & Excellent \cite{doi:10.1080/10447310802205776} \\
\hline
Support Application & 77.04  & Excellent \cite{doi:10.1080/10447310802205776} \\
\hline
\end{tabularx}
\end{table}

Both applications received an "Excellent" score in terms of usability scale based on the interpretation of SUS Results proposed by Bangor at al. \cite{doi:10.1080/10447310802205776}. Overall score results of both applications are shown in \cref{table:SUSQR-OVERALL}.

\subsection{NASA TLX}

The evaluation of the NASA Task Load Index results is based on the work of S. G. Hart \cite{HART1988139}. Nominal descriptions of the values are referenced based on \cref{table:NASA-TLX_SCORE_INTERPRETATION}.

\begin{table}[h]
\centering 
\caption{Interpretation of NASA TLX scores \cite{HART1988139}}
\label{table:NASA-TLX_SCORE_INTERPRETATION}
\begin{tabular}{| l | c |}
\hline
\textbf{Workload} & \textbf{Value} \\
\hline
Low & 0-9 \\
\hline
Medium & 10-29 \\
\hline
Somewhat high & 30-49 \\
\hline
High & 50-79 \\
\hline
Very high & 80-100 \\
\hline
\end{tabular}
\end{table}

The mental demand was perceived to be somewhat high, suggesting that cognitive processing played a significant role in the task. Physical demand was rated as medium, indicating that participants experienced moderate physical exertion. Temporal demand was also rated as medium, suggesting that participants felt a moderate sense of time pressure or time-related stress.

Despite these demands, participants reported very high levels of performance, indicating that they felt confident in their ability to effectively complete the task. Effort expended during the task was perceived to be medium, suggesting a moderate level of exertion or workload required. Additionally, participants reported a medium level of frustration, indicating some degree of annoyance or difficulty encountered during the task, but not to an extreme extent.

\begin{figure}[h]    
    \centering
    \includegraphics[width=0.8\textwidth]{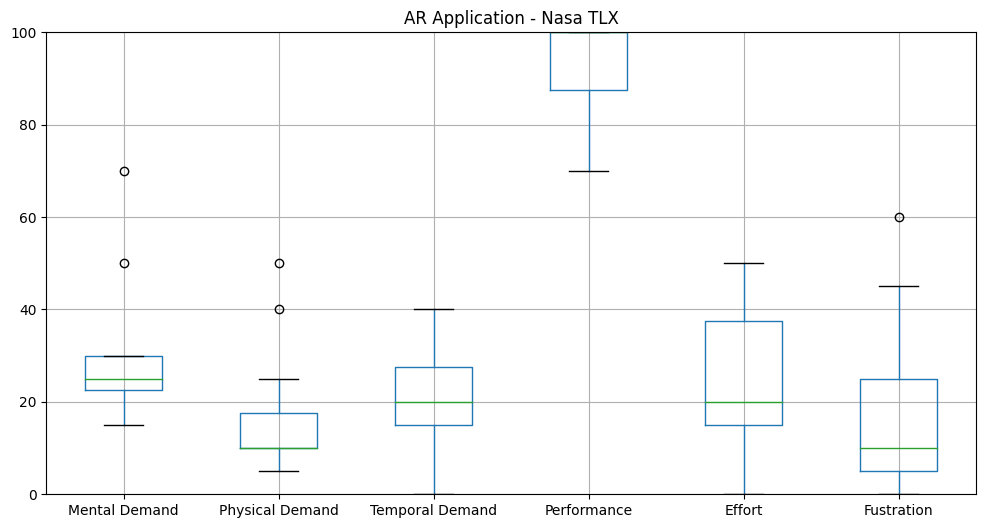}
    \caption{NASA TLX - AR APPLICATION RESULTS}
    \label{fig:NASA-TLX-AR}
\end{figure}

The average results are presented in \cref{table:AR-NASA-TABLE}, and the complete collected data is depicted in \cref{fig:NASA-TLX-AR}.

\begin{table}[h]
\centering
\caption{AR Application - NASA TLX Average Scores}
\label{table:AR-NASA-TABLE}
\begin{tabular}{| l | c |}
\hline
\textbf{Category} & \textbf{Average Score}  \\
\hline
Mental Demand & 31\% \\
\hline
Physical Demand & 17\% \\
\hline
Temporal Demand & 20\% \\
\hline
Performance & 93\% \\
\hline
Effort & 26\% \\
\hline
Frustration & 18\% \\
\hline
\end{tabular}
\end{table}

The mental demand was perceived to be somewhat high, indicating that cognitive processing played a significant role in the task. However, physical demand was rated as low, suggesting that participants experienced minimal physical exertion.

Temporal demand was rated as medium, suggesting that participants felt a moderate sense of time pressure or time-related stress. Despite these demands, participants reported very high levels of performance, indicating that they felt confident in their ability to effectively complete the task.

Effort expended during the task was perceived to be medium, suggesting a moderate level of exertion or workload required. Additionally, participants reported a medium level of frustration, indicating some degree of annoyance or difficulty encountered during the task, but not to an extreme extent.

\begin{figure}[h]    
    \centering
    \includegraphics[width=0.8\textwidth]{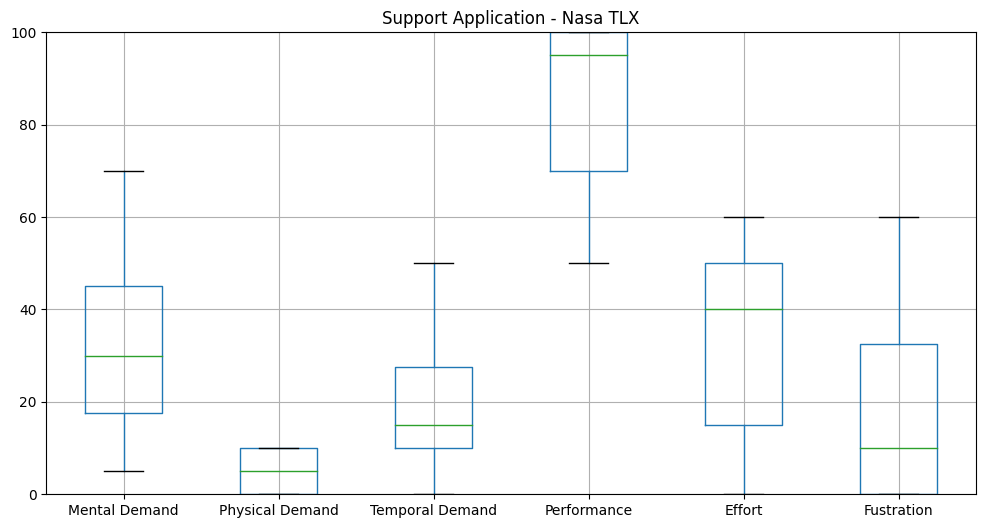}
    \caption{NASA TLX - SUPPORT APPLICATION RESULTS}
    \label{fig:NASA-TLX-WEB}
\end{figure}

The average results are presented in \cref{table:SUPPOR-NASA-TABLE}, and the complete collected data is depicted in \cref{fig:NASA-TLX-WEB}.

\begin{table}[h]
\centering
\caption{Support Application - NASA TLX Average Scores}
\label{table:SUPPOR-NASA-TABLE}
\begin{tabular}{| l | c |}
\hline
\textbf{Category} & \textbf{Average Score}  \\
\hline
Mental Demand & 31\% \\
\hline
Physical Demand & 5\% \\
\hline
Temporal Demand & 18\% \\
\hline
Performance & 85\% \\
\hline
Effort & 35\% \\
\hline
Frustration & 20\% \\
\hline
\end{tabular}
\end{table}

Comparing the results of the two applications, the data shows that the AR application requires higher physical effort from the user. The additional weight of the HMD device, worn by the maintenance personnel, affects the user's comfort.
At the same time, the Support Application, based on the participant feedback, requires more effort and is less performant. This result was expected, as the support application allows more functionality to delegate tasks from the maintenance personnel.

\subsection{Additional Qualitative Results}

Additional qualitative data was collected regarding aspects of the solutions that are not covered by the standard NASA-TLX or SUS methods. These were conducted using a 5-point Likert scale, where we consider 1 as the lowest value and 5 as the highest value.

Overall, the AR Application has been perceived positively, as depicted in \cref{table:AR-QUESTIONER}. While there is a consensus about the overall usefulness of the AR application, receiving a 4.5 regarding improving productivity, consideration of using such technology, and the benefit of the voice communication capabilities, the "Action Card" feature received the highest score with 4.8. Participants found it useful for information to be displayed in a variety of ways, that could be adapted based on the type of information the situation needs to be displayed. 

However, the option of the support personnel to change the information displayed by the "Action Card" could be considered a useful feature, participants rated this slightly lower than expected. The scenarios the lab environment allows are less complex than real-world scenarios, giving participants more freedom to interact with virtual objects. Further experimentation with real-world scenarios might show more benefit to the feature, as complex maintenance processes might require both hands, therefore putting more restrictions on the maintenance personnel to interact with the virtual objects.

The least positively viewed feature was the 3D representation of the Support personnel in the work environment. Participants, although they agreed that the representation of the support personnel could be beneficial, expressed that the implementation would have a higher impact with further interactivity. Similarly, to the manipulation of the "Action Cards", the capability of the support personnel to point at certain locations in the real-world environment could not be fully realized in the lab environment, due to the confined space. Further real-life experimentation would be beneficial to improve our understanding of these features.

\begin{table}[h]
\centering
\caption{AR Application - Additional Qualitative Results}
\label{table:AR-QUESTIONER}
\begin{tabularx}{\columnwidth}{| X | c |}
\hline
\textbf{Statement} & \textbf{Average Result}  \\
\hline
I would consider using AR technology for similar scenarios & 4.5 \\
\hline
I found the voice communication helpful & 4.5 \\
\hline
I believe the 3D representation of the support personnel can be beneficial & 3.8 \\
\hline
The "Action cards" are useful for displaying information & 4.8 \\
\hline
It's beneficial that "Action cards" could be controlled by the support personnel & 4.1 \\
\hline
I would consider this method to have a direct connection with other people to improve productivity & 4.5 \\
\hline
\end{tabularx}
\end{table}

\begin{table}[h]
\centering
\caption{Support Application - Additional Qualitative Results}
\label{table:SUPPORT-QUESTIONER}
\begin{tabularx}{\columnwidth}{| X | c |}
\hline
\textbf{Statement} & \textbf{Average Result}  \\
\hline
I found the voice communication really helpful & 4.6 \\
\hline
The "Video Feed" made communication with the Maintenance personnel easier. & 4.7 \\
\hline
The "3D View" helped with communication in the environment. & 3.4 \\
\hline
I would consider this method to have a direct connection with other people to improve productivity & 4.4 \\
\hline
\end{tabularx}
\end{table}

Similarly, to the AR application, the Support application is also considered overall a beneficial enhancement as a tool for maintenance processes. As \cref{table:SUPPORT-QUESTIONER} shows, both voice communication and the "video feed" are viewed as a useful feature for assisting the maintenance personnel, with the "video feed" being slightly more preferred.

Considering, that the 3D View, which creates a 3D digital representation of the real-world working environment, is incapable of recording color information, resulting in only a greyscale 3D mesh, most of the participants operating the support application deemed the least useful for assessing the environment. Additionally, to this visual limitation, the lab environment, where the testing took place, might not reflect the extent of the capabilities of the technology, due to the confined space. Further improvements and testing in a larger environment might yield more favorable results.

\section{Conclusion}

We have presented a comprehensive exploration of the integration of AR technology into industrial maintenance practices, with a focus on enhancing user experience and efficiency. The manufacturing industry's evolution has underscored the critical role of maintenance in ensuring production efficiency and avoiding costly disruptions. Traditional maintenance strategies, predominantly reactive and preventive, are gradually giving way to proactive approaches, recognizing maintenance as a strategic factor and profit contributor in industrial systems.

Our proposed solution introduces an intuitive and modular maintenance system rooted in Industry 5.0 principles, leveraging AR interfaces to provide guided decision support for maintenance technicians. By superimposing virtual data onto the real world, including images, text instructions, and 3D models, AR technology facilitates complex maintenance tasks, leading to improved productivity and reduced downtime. The integration of HMD Devices and support interfaces creates a seamless human-machine interaction environment in simulated industrial settings.

Throughout the paper, we have discussed related work in the field, including UI systems designed to guide technicians through maintenance tasks using AR. We have also examined practical implementations and validations of similar frameworks in both laboratory-based and real-life industrial scenarios. These insights provide valuable context and comparison points for our proposed solution's development and evaluation.

Our methodology encompasses the development of AR maintenance and support applications, detailed procedures for evaluation, and the utilization of qualitative metrics such as the NASA-TLX and the SUS to assess usability, performance, and workload implications. By conducting experiments in simulated environments, we aimed to provide insights into the effectiveness of our solution while ensuring participant safety and comfort.

The results of our evaluations have shown promising usability and performance metrics, with participants expressing positive feedback regarding system usability, ease of use, and integration of functions. While some challenges such as physical exertion and minor inconsistencies have been identified, the overall user experience remains favorable. Furthermore, the comparison between the AR maintenance application and the support application revealed nuanced differences in user perceptions, highlighting areas for improvement and optimization in future iterations.

\section{Acknowledgment}

%This project has received funding from the European Union’s Horizon Europe Research and Innovation programme under grant agreement No. 101070181.

This work was funded by UK Research and Innovation (UKRI) under the UK government’s Horizon Europe funding guarantee [grant number 10047653] and funded by the European Union [under EC Horizon Europe grant agreement number 101070181 (TALON)].

\bibliography{bib.bib}
\bibliographystyle{ieeetr}

\end{document}